\documentclass[%
superscriptaddress,
frontmatterverbose,  
preprint,
preprintnumbers,
nofootinbib,
amsmath,
amssymb, 
aps, 
prl,
floatfix 
]{revtex4-1}

\usepackage{amsfonts}       
\usepackage{amsmath}       
\usepackage{amssymb}
\usepackage{nicefrac} 
\usepackage{graphicx}
\usepackage{dcolumn}
\usepackage{bm}
\usepackage{hyperref}
\usepackage{xcolor}  
\usepackage{array}
\usepackage[export]{adjustbox}
\usepackage{multirow}
\usepackage[figure,table]{totalcount}
\usepackage{verbatim}

\usepackage{changes} 

\usepackage{xifthen} 

\usepackage{todonotes}

\newcommand{\avg}[1]{\left\langle#1\right\rangle} 

\makeatletter
\setremarkmarkup{\todo[color=Changes@Color#1!20,size=\scriptsize]{#1: #2}}
\makeatother

\definechangesauthor[name={Mauricio}, color=red]{M} 
\definechangesauthor[name={Osame}, color=blue]{O} 

\begin{document}

\title{Synaptic balance due to homeostatically self-organized quasi-critical dynamics}

\author{Mauricio Girardi-Schappo$^*$}
\affiliation{Universidade de S\~ao Paulo, FFCLRP, Departamento de F\'isica, Ribeir\~ao Preto, SP, 14040-901, Brazil}

\author{Ludmila Brochini}
\affiliation{Universidade de S\~ao Paulo, Instituto de Matem\'atica e Estat\'istica, S\~ao Paulo, SP, 05508-090, Brazil}

\author{Ariadne A. Costa}
\affiliation{Universidade Federal de Goi\'as - Regional Jata\'i, Unidade Acad\^emica Especial de Ci\^encias Exatas, Jata\'i, GO, 75801-615, Brazil}

\author{Tawan T. A. Carvalho}
\affiliation{Universidade Federal de Pernambuco, Departamento de F\'isica, Recife, PE, 50670-901, Brazil}

\author{Osame Kinouchi}
\email{osame@ffclrp.usp.br}

\thanks{girardi.s@gmail.com; These authors contributed equally to this work.}
\affiliation{Universidade de S\~ao Paulo, FFCLRP, Departamento de F\'isica, Ribeir\~ao Preto, SP, 14040-901, Brazil}

\date{\today}

\begin{abstract}
Recent experiments suggested that homeostatic regulation of synaptic balance
leads the visual system to recover and maintain a regime of power-law avalanches.
Here we study an excitatory/inhibitory (E/I) mean-field neuronal network
that has a critical point with power-law avalanches and synaptic balance.
When short term depression in inhibitory synapses and firing threshold adaptation
are added, the system hovers around the critical point.
This homeostatically self-organized quasi-critical (SOqC) dynamics generates E/I synaptic
current cancellation in fast time scales, causing
fluctuation-driven asynchronous-irregular (AI) firing.
We present the full phase diagram of the model without adaptation
varying external input versus synaptic coupling.
This system has a rich dynamical repertoire of
spiking patterns: synchronous regular (SR), 
asynchronous regular (AR), synchronous irregular (SI), slow oscillations (SO) and AI.
It also presents dynamic balance of synaptic currents, since inhibitory currents
try and compensate excitatory currents over time, resulting in
both of them scaling linearly with external input.
Our model thus unifies two different perspectives on cortical
spontaneous activity: both critical avalanches 
and fluctuation-driven AI firing arise from SOqC
homeostatic adaptation, and are indeed 
two sides of the same coin.
\\
$ $\\
\textbf{DOI:} \href{https://link.aps.org/doi/10.1103/PhysRevResearch.2.012042}{10.1103/PhysRevResearch.2.012042}

%
%
%
\end{abstract}


\maketitle

Experimental and theoretical evidence suggests
that spontaneous cortical activity happens in the form of asynchronous 
irregular firing patterns (AI). This could be generated by the balance 
of excitatory/inhibitory (E/I) synaptic currents entering
individual neurons (see~\cite{Deneve2016,Ahmadian2019}):
inhibition has to nearly compensate excitation, such that cells remain
near their firing threshold and fire sporadically, generating a fluctuation-driven
regime~\citep{Ahmadian2019}.
These firings may be organized in avalanches of action potentials
that spread throughout the cortex.
Critical avalanches are known to enable the propagation of
fluctuations through local interactions due to long-range
spatiotemporal correlations~\citep{Odor2004}, generating optimized
processing and functional
features~\citep{Haldeman2005,Kinouchi2006,Shew2009,Girardi2016,Mosqueiro2013}.

Two important issues remain: (i) how to self-organize 
a neuronal network close to
a critical point, and (ii) could a network
display an AI firing pattern through this self-organization?
Concerning the first point, it has been shown that simple
local homeostatic mechanisms, such as dynamical
synapses~\citep{Levina2007,Bonachela2010,Peng2013,Costa2015,Campos2017}
and dynamical neuronal gains~\citep{Brochini2016,Costa2017,Kinouchi2019},
are sufficient to drive
networks towards the so-called Self-Organized quasi-Critical
state (SOqC as defined by
Bonachela \& Mu{\~n}oz~\cite{Bonachela2009,Bonachela2010}). 
Particularly, our model requires two independent
homeostatic mechanisms to generate the SOqC dynamics:
plasticity in the inhibitory synapses~\citep{Lambert1994} and 
adaptive firing thresholds~\citep{Clifford2007}.

As for the second point, we will show that 
our homeostatic mechanisms for SOqC generate a near cancellation 
of excitatory/inhibitory (E/I) synaptic currents that produces
a fluctuation-driven AI regime.
Therefore, AI is a direct consequence of the hovering around a critical point
where the system displays quasi-critical power-law avalanches.
Indeed, recent experiments show homeostatic
regulation of network activity close to a critical 
state happening most probably through the adaptation
of inhibitory synapses~\cite{Ma2019}.

There have been attempts to model E/I networks in the context of
criticality~\citep{Poil2012,Lombardi2012,Lombardi2017,DallaPorta2019}.
However, none of these models have shown that neuronal avalanches
with the correct exponents
arise when E/I synaptic currents cancel each other. 
Also, none of these models show that synaptic currents balance each other
in the vicinity of a critical point. Not only the SOqC dynamics proposed here
does that, but it also generates activity where avalanches and AI spiking coexist.

Without SOqC, we have a static system
presenting the typical synchronicity states of
E/I networks exemplified by Brunel's model~\citep{Brunel2000}:
synchronous regular (SR), asynchronous regular (AR),
synchronous irregular (SI) and asynchronous irregular (AI).
This system has a directed percolation (DP) critical point with
power-law avalanches, and dynamic balance of E/I currents, since
inhibitory inputs follow excitatory ones over time.
Even though the E/I neuron ratio is 80\%:20\%
from cortical data~\citep{Somogyi1998}, our model predicts that
the ratio of coupling strengths of inhibitory to excitatory synapses does
not need to be 4:1 to achieve the critical balanced state.

We first define both the static and
the adaptive versions of the model. Then, we make a 
mean-field calculation obtaining the
critical exponents and phase diagrams, 
and discuss the dynamic states of the
static network. Finally, we add SOqC homeostatic adaptation
and observe the hovering around the critical balanced point that displays near
cancellation of E/I currents and fluctuation-driven AI activity.

\section{The Model}
We use discrete-time 
stochastic integrate-and-fire 
neurons~\citep{Gerstner1992, Galves2013,Brochini2016}.
A Boolean variable denotes if a neuron fires ($X[t]=1$) or not
($X[t] = 0$) at time $t$. The membrane
potentials of neurons in $E$ and $I$ populations evolve as:
\begin{align}
\label{VE}
V_i^E[t+1] = \Bigg[ \mu V_i^E[t] + I_i^{(E)}[t]
+\dfrac{1}{N} \sum_{j=1}^{N_E} W_{ij}^{EE} X_j^E[t]
-\dfrac{1}{N} \sum_{j=1}^{N_I} W_{ij}^{EI} X_j^I[t] \Bigg]
\bigg(1-X_i^E[t] \bigg),
\\
\label{VI}
V_i^I[t+1] = \Bigg[\mu V_i^I[t] + I_i^{(I)}[t]
+\dfrac{1}{N} \sum_{j=1}^{N_E} W_{ij}^{IE} X_j^E[t]
-\dfrac{1}{N} \sum_{j=1}^{N_I} W_{ij}^{II} X_j^I[t] 
\Bigg] \bigg(1-X_i^I[t]\bigg),
\end{align}
where $N = N_E + N_I$ is the total number of neurons,
$\mu$ is a leakage parameter and $I_i^{(E)/(I)}[t]$ 
are external inputs over $E$ and $I$ populations, respectively. 
The second index in $W_{ij}^{ab}$, with $a,b \in \{E,I\}$,
refers always to the presynaptic neuron. 
All the $W$'s are positive
(inhibition is explicitly given by the minus).
The term $(1-X_i[t])$ resets the voltage to zero after
a spike, resulting in one time step of refractoriness.
Our network is fully connected with $K = N-1$ neighbors.

\begin{figure}[tp]
\begin{center}
\centerline{
\includegraphics[width=0.98\textwidth]{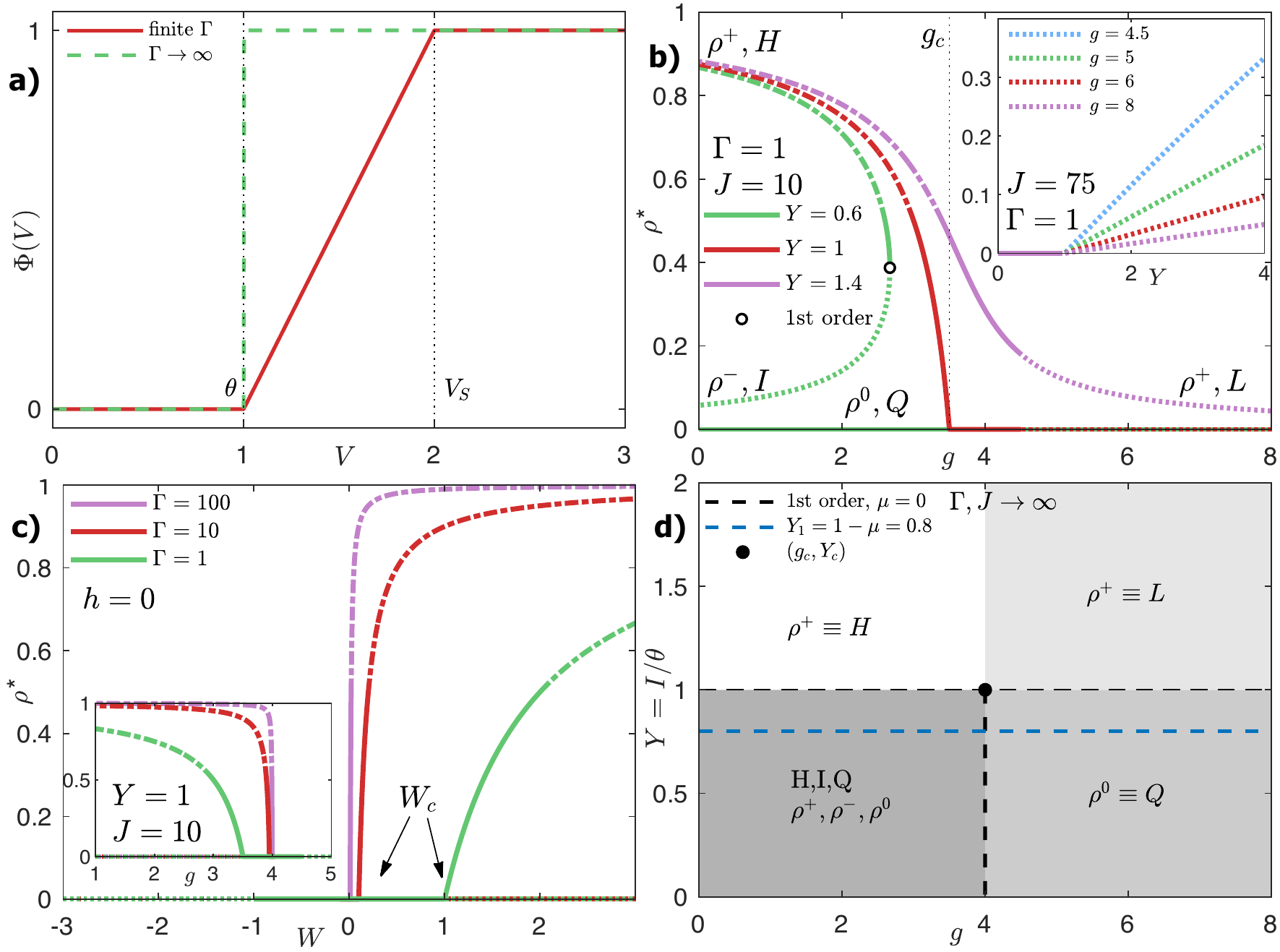}}
\caption{{\bf Firing rate function and phase transitions.}
{\bf a,} Solid: soft firing threshold ($\Gamma=1$),
dashed: hard threshold ($\Gamma \rightarrow \infty$).
{\bf b,} Order parameter \textit{vs.} $g$ (inset: \textit{vs.} $Y$),
highlighting the activity states high (H), low (L),
intermediary (I, unstable, from a fold bifurcation)
and quiescent (Q $\equiv \rho^0=0$).
{\bf c,} Order parameter $\rho^*$ \textit{vs.} $W$ for $h=0$
(inset: \text{vs.} $g$ for $Y=1$);
notice as the critical point shifts away from $W_c=0$ ($g_c=4$)
as $\Gamma$ decreases ($W_c = 1/\Gamma$).
{\bf b and c,} Dot-dashed lines are marginally stable cycle-2 attractors
(SR state). Dotted lines in the $\rho^+$(L) branch are cyclic attractors
of the network (quasi-cycle-2 SI states). Density $\rho$ is given
by Eqs.~\eqref{FOT} and~\eqref{FOT2}.
{\bf d,} Phase diagram in the balanced notation
$(g,Y)$ plane for the hard threshold neurons.
The critical point lies at $(W_c = 0, h_c = 0)$ or 
$(g_c = 4, Y_c = 1)$ [bullet, Eq.~\eqref{gc}].
The Q phase loses stability at the horizontal dashed
line $Y_c =1-\mu$ (or $h_c = 0$); $\mu=0$ (black and thin dashed line)
and $\mu = 0.2$ (blue dashed line).
This diagram should be compared to Fig.~1A of Ref.~\cite{Brunel2000}.}
\label{PHI}
\end{center}
\end{figure}

The individual neurons fire following a piece-wise linear  
probability function (see Fig.~\ref{PHI}a):
\begin{equation}
P\left(X=1|V\right) \equiv \Phi(V)
= (V-\theta)\:\Gamma\:\Theta(V-\theta)\:
\Theta(V_S - V) + \Theta(V-V_S),
\label{phifunc}
\end{equation}
where $\Gamma$ is the neuronal firing gain, 
$\theta$ is a firing threshold, $V_S = \theta + 1/\Gamma$
is the saturation potential and
$\Theta(x)$ is the Heaviside function.
The firing probability $\Phi(V)$ captures the effects of membrane noises,
inducing stochastic spiking.
The limit $\Gamma \rightarrow \infty$ reduces to the
leaky integrate-and-fire (LIF)
neuron with hard threshold $V_S=\theta$.

\section{Order and control parameters}
We assume that the
synaptic weights have finite variance (are self-averaging),
approximating them by their mean values
$W^{ab} =\avg{W_{ij}^{ab}}$ (for all the $a,b \in \{E,I\}$).
We also define the firing densities
(the fraction of active sites) $\rho_E[t] = 1/N_E \sum_j X_j^E[t]$ and
$\rho_I[t] = 1/N_I \sum_j X_j^I[t]$. 
The fractions of excitatory and inhibitory
neurons are $p = N_E/N$ and $q = 1-p = N_I/N$, respectively.
Finally, we consider only the case with a stationary
average external input $I = \avg{I_i[t]}$ with finite variance over both populations.

We introduce the synaptic
balance parameter $g$ by letting the synaptic
weights obey $W^{EE}=W^{IE}=J$, and
$W^{II}=W^{EI}=gJ$ (Brunel's model A~\cite{Brunel2000}).
This is not a necessary assumption, but it reduces
Eqs.~\eqref{VE} and~\eqref{VI} to a single iterative map
that is equal for both $E/I$ populations:
\begin{equation}
V_i[t+1] = \bigg[ \mu V_i[t] + I +
 p  J \rho_E[t]
- q  g J \rho_I[t]  \bigg] \bigg( 1 - X_i[t]\bigg)\:,
\label{VAvg}
\end{equation}
where we may omit the $E/I$ superscripts.
Letting the excitatory synaptic current be\footnote{Not to be confused with external input over the excitatory population $I_i^{(E)}[t]$ in Eq.~\eqref{VE}.}
$I^E[t]=pJ\rho_E[t]$
and the inhibitory be\footnote{Not to be confused with external input over the inhibitory population $I_i^{(I)}[t]$ in Eq.~\eqref{VI}.}
$I^I[t]=-qgJ\rho_I[t]$,
we define the average net synaptic current,
\begin{equation}
\Delta I^{E/I}=I^E+I^I=pJ\rho_E - qgJ\rho_I=W\rho\:,
\label{netsyn}
\end{equation}
where we used $\rho_E=\rho_I=\rho$ (from the constraints imposed on the synaptic weights)
and defined $W=(p-qg)J$ as our first control parameter.
This holds because, after a neuron spikes, the 
voltage reset erases initial conditions and
the voltages for both $E/I$ populations
evolve  following  Eq.~\eqref{VAvg}.
The firing density $\rho$ is our order parameter,
equivalent to the network firing frequency
$\nu_0$ of Brunel's model~\cite{Brunel2000}.

Consider the stationary state $(1-\mu)V^* = I$ for
 $\rho = 0$ in Eq.~\eqref{VAvg}. When
 $V^* = \theta$, $\Phi(V^*) = 0$, so
 we have $\rho>0$ for $I>(1-\mu)\theta$. Thus, we
define the external field  $h=I-\theta(1-\mu)$ as
the average suprathreshold external current. The $h$ variable
is our second control parameter.
The parameters $(W,h)$ are usual for Statistical Physics.
By introducing the external current ratio, $Y=I/\theta$,
we may switch from $(W,h)$ to describe the system in
the balanced notation $(g,Y)$ by using
$g = p/q - W/(qJ)$ and
$Y = (h/\theta)+1-\mu$.

\section{Homeostatic mechanisms}
To obtain a quasi-critical balanced state without fine tuning,
we introduce two independent homeostatic biological mechanisms:
inhibition depression~\citep{Ma2019}
and firing threshold adaptation~\citep{Benda2003}.
We use the Levina-Hermann-Geisel 
short-term plasticity for the synaptic weights~\citep{Levina2007}:
\begin{equation}
W^{II/EI}_{ij}[t+1] =
W^{II/EI}_{ij}[t] + \dfrac{1}{\tau_W} \left(A-W^{II/EI}_{ij}[t] \right)
- u_W W^{II/EI}_{ij}[t] X_j^I[t] \:,
\label{WII}
\end{equation}
where $\tau_W$ is a (large) recovery time, $A$ is the synaptic baseline and 
$u_W$ is the fraction of the synaptic 
strength depressed when a presynaptic neuron fires. 
This dynamic generates homeostatic tuning because $g$ is then
$g[t] = \avg{W_{ij}^{EI/II}[t]}/J$ in Eq.~\eqref{VAvg}.
The $\avg{.}$ bracket is an average over neurons $i$ and $j$.

To self-organize towards zero-field 
$h_c = I -(1-\mu)\:\theta = 0$ 
or $Y_c = I/\theta = 1-\mu$,
we add threshold adaptation:
\begin{equation}
\theta_i[t+1] =\theta_i[t] - \dfrac{1}{\tau_\theta}
\theta_i[t]   + u_\theta \theta_i[t] X_i[t]\:,
\end{equation}
where the parameter $u_{\theta}$ is the 
fractional increase in the
neuron threshold after it fires, and 
$\tau_{\theta}$ is a recovery
time scale. This dynamic is inspired by the biological
mechanism of firing rate adaptation~\citep{Benda2003}.
It enters the model through Eq.~\eqref{phifunc},
changing $\theta$ to $\theta[t]=\avg{\theta_i[t]}$.

\section{Mean-field calculations} 
We consider only the $\mu=0$, since $\mu>0$
does not present any new phenomenology
(although it admits numerical solutions 
and analytic approximations close the critical
point~\citep{Brochini2016,Costa2017}).
For this case, the stationary voltage distribution has
only two delta peaks,
$P_t(V) = \rho[t]
\delta\left(V\right)+\left(1-\rho[t]\right)\,\delta
(V - V[t])$, and the number of active sites is the average of
 $\Phi(V)$ over $V$~\citep{Brochini2016,Kinouchi2019},
\begin{equation}
\rho[t+1] = \int \Phi(V) \:P_t(V) \: dV\:,
\label{rhoavg}
\end{equation}
with $V[t]$ given by Eq.~\eqref{VAvg},
resulting in:
\begin{equation}
\label{rhosfinal}
\rho[t+1] = \left(1-\rho[t]\right)\Gamma
\left( W \rho[t]+ h \right)\Theta(W \rho[t]+ h)\:.
\end{equation}
This map has, in principle, three fixed points.
For $h\leq 0$, there is a quiescent solution
$\rho^0=0$ (also called the Q state) since
the Heaviside $\Theta(x)$ function is zero 
in the right hand side in Eq.~\eqref{rhosfinal}.

The active states are the two other fixed points of the 
firing density Eq.~\eqref{rhosfinal},
given by:
\begin{equation}
\Gamma W \rho^2 + \left(1 + \Gamma h - \Gamma W
\right) \rho - \Gamma h = 0 \:,  \label{eqrhomu0} 
\end{equation}
with solutions:
\begin{equation}
 \rho^\pm  =  \dfrac{\Gamma W -\Gamma h - 1}{2\Gamma W}
 \pm\dfrac{\sqrt{(\Gamma W-\Gamma h-1)^2 + 4 \Gamma^2 W  h}}{2\Gamma W}.
\label{FOT} 
\end{equation} 
For $h > 0$ ($Y > 1$), there is a single solution $\rho^+$ (corresponding to
high activity H and low activity L) because
$\rho^-<0$. For $h < 0$ ($Y < 1$), we have a positive
but unstable branch $\rho^-$ (the intermediary solution I) that separates the
stable branch $\rho^+$ (H) from the absorbing state $\rho^0$ (Q),
see Fig.~\ref{PHI}b. When $h=0$ ($Y=1$), the unstable branch vanishes
into a critical point with $W=W_c=1/\Gamma$ [$g=g_c$, Eq.~\eqref{gc}].

\section{Critical exponents}
For zero-field, Eq.~\eqref{eqrhomu0}
yields $\rho^0=0$  (the absorbing quiescent phase, Q),
stable for $W<W_c \equiv 1/\Gamma$ 
and an active state:
\begin{equation}  \label{trans}
\rho^* = \frac{\Gamma W-1}{\Gamma W} = \frac{W - W_c}{W}\sim(W-W_c)^{\beta}\:,
\end{equation}
with $\beta=1$, stable for $W > W_c = 1/\Gamma$. The field exponent
is obtained by isolating $h$ from Eq.~\eqref{eqrhomu0} and 
expanding for small $\rho$ (due to small external $h$) with $W=W_c$,
resulting in $\rho^*\sim(h/W_c)^{1/\delta_h}$ with $\delta_h=2$.
The exponent of the susceptibility,
$\chi=\partial\rho/\partial h\sim|W-W_c|^{-\gamma}$,
using $\Gamma=1/W_c$, is $\gamma=1$. 

These exponents pertain to
the mean-field directed percolation (DP) universality
class~\citep{Dickman1999,Lubeck2004,Girardi2019},
the framework that has been proposed to describe neuronal 
avalanches~\citep{Chialvo2010,Munoz2018}.
The variance of the network activity is ${\rm
Var}(\rho)\sim|W-W_c|^{-\gamma^\prime}$
with $\gamma^\prime=0$~\citep{Dickman1999}.
This explains the jump in the coefficient of variation of $\rho$
observed by Brunel~\citep{Brunel2000}.

In the balanced notation, $h=h_c=0$ is the same as
$Y_c = (h_c/\theta)+1-\mu = 1$ 
(recalling that $\mu=0$ and $\theta=1$).
The equivalent of $W_c=1/\Gamma$ is given by
\begin{equation}
\label{gc}
g_c = \dfrac{p}{q} - \dfrac{W_c}{q J} = 4 - \dfrac{5}{\Gamma J} \:,
\end{equation}
where the usual cortical
estimates $p=80\%$ and $q=20\%$ were used~\citep{Somogyi1998}.
This generalizes the usual condition $g_c\approx 4$:
if neurons have a soft threshold (finite $\Gamma$)
or the synapses are weak (finite $J$), 
the critical balance point
shifts towards lower values of $g$ (Fig.~\ref{PHI}c).
The phase diagram for large $\Gamma J$ (\textit{i.e.}
hard threshold LIF neurons) is shown in Fig.~\ref{PHI}d, and
matches exactly the one obtained by Brunel~\citep{Brunel2000}.

\begin{figure*}[tp]
\begin{center}
\centerline{
\includegraphics[width=0.98\textwidth]{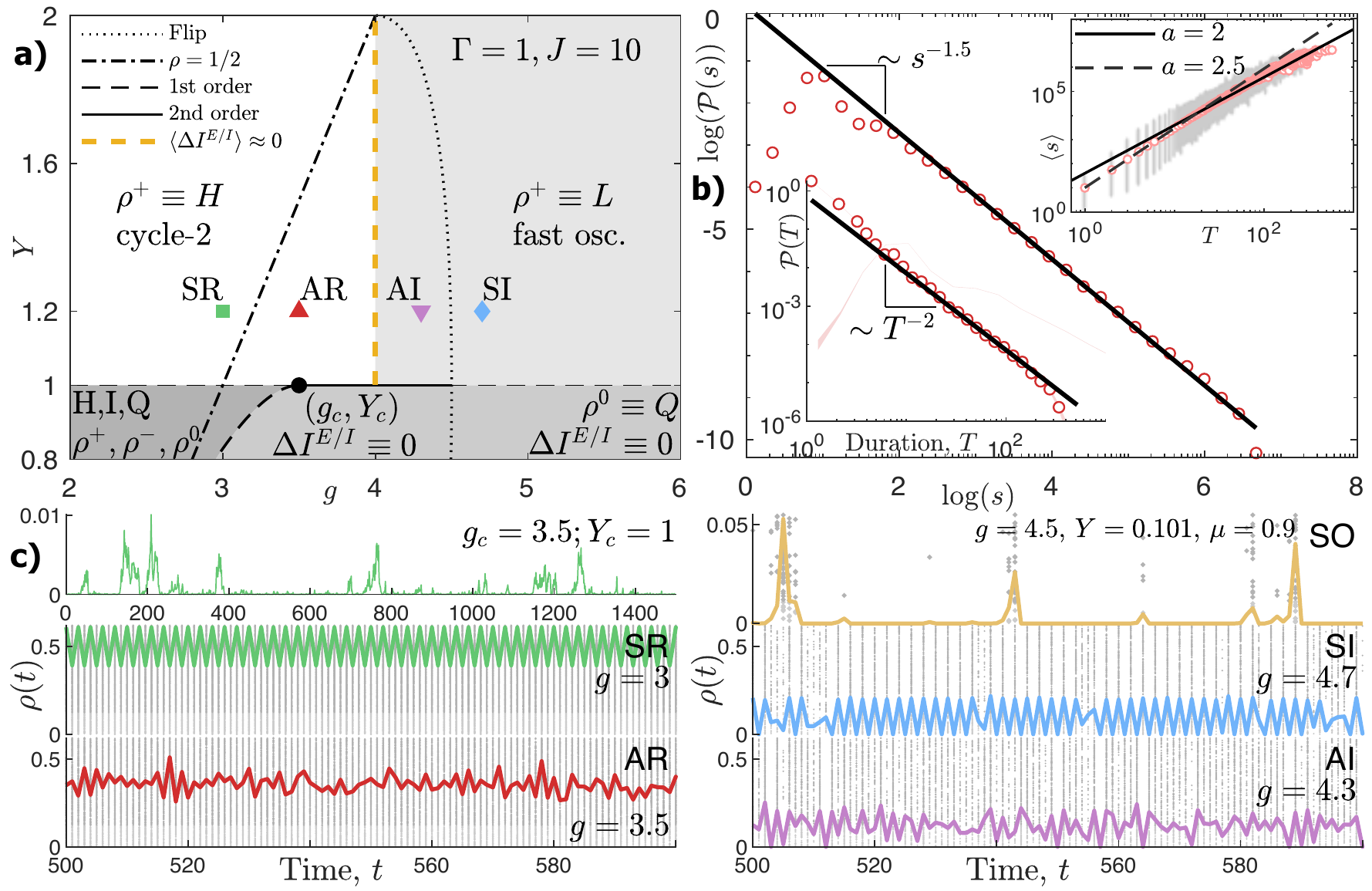}}
\caption{{\bf Avalanches and firing patterns.}
{\bf a,} Phase diagram for $\mu=0$, and $\Gamma J = 10$;
a critical line starts at $g_c=3.5$, see Eq.~\eqref{gc}, for $Y=1$.
The critical point, the subcritical region with $g>g_{\rm Fold};\,Y\leq1$,
and the supercritical region $g=4;\,Y>1$ have balanced synaptic currents,
such that the net current is $\Delta I^{E/I}=I^E+I^I \approx 0$.
At $Y=1.2$, from left to right: SR/cycle-2 ($g=3$),
AR/High ($g=3.5$), AI/Low ($g=4.3$), and SI/fast oscillations ($g=4.7$).
SR and AR are separated by a bifurcation 
tho cycle-2 due to the refractory
period; SI and AI are separated by a flip bifurcation.
{\bf b,} Distribution of avalanche sizes (main plot, $\tau=1.5$)
and duration (bottom inset, $\tau_t=2$) at the critical point;
Top inset: size and duration scaling law $\avg{s}\sim T^a$
has a crossover with $a=2.5$ for small avalanches (a finite-size effect)
and $a=2$ for the rest of the data.
{\bf c,} Network simulation results ($N=10^6$ neurons), $\rho[t]$,
for the points in panel \textbf{a}. From the top left to the bottom right
panel: critical point absorbing-state avalanches (peaks); SR, AR, SO
(slow waves for $Y\gtrsim Y_c=1-\mu$,
$\mu=0.9$, $Y=0.101$), SI, and AI. The background shows the raster plot
of 1,000 randomly selected neurons.}
\label{aval}
\end{center}
\end{figure*}

\section{Synaptic currents of the static model}
We can write Eq.~\eqref{FOT} in the balanced notation
by letting $h = (Y-1)\theta$ and $W=(p-qg)J$ [see Fig.~\ref{PHI}b]:
\begin{equation}
\rho^\pm =\dfrac{1}{2}+\dfrac{\rho_1}{2\Gamma\theta(Y-1)}+\dfrac{\rho_1}{2}  
\pm \sqrt{\left(\dfrac{1}{2}+\dfrac{\rho_1}{2\Gamma\theta(Y-1)}+\dfrac{\rho_1}{2}\right)^2
-\rho_1}\:,
\label{FOT2}
\end{equation}
where $\rho_1=(Y-1)\theta/[Jq(g-p/q)]$ 
is the first order expansion
of Eq.~\eqref{FOT2}.

The synaptic currents
are balanced if the net synaptic current from Eq.~\eqref{syncurr} is zero,
$\Delta I^{E/I}=W\rho=0$, such that either $W=(p-qg)J=0$ (\textit{i.e.},
$g=g_{\rm bal}=p/q$ for $Y>1$), or $\rho=0$ (\textit{i.e.}, the quiescent solution of the
subcritical and critical states, $g\geq g_c$ and $Y\leq1$).
For $g\neq p/q$, the synaptic currents scale linearly with
the external input. We can see that by expanding Eq.~\eqref{FOT2}
for small $Y$, giving $\rho\approx\rho_1$:
\begin{align}
    I^E=pJ\rho_1&=\dfrac{p/q}{g-p/q}(Y-1)\\
    I^I=-qgJ\rho_1&=-\dfrac{g}{g-p/q}(Y-1)\:.
\end{align}
The variable $\rho\approx\rho_1=I^E/(pJ)$ 
is shown in the inset of
Fig.~\ref{PHI}b. These currents saturate for large enough $\Gamma J$.
This linear scaling highlights the dynamic balance of synaptic inputs,
as inhibition tracks excitation over time~\cite{Van1996,Brunel2000}.

\section{Phase diagram} 
The soft threshold  neurons' 
phase diagram is shown in
Fig.~\ref{aval}a.
The curves are bifurcations of the stable fixed point $\rho^+$ in Eq.~\eqref{FOT2}:
(i) a fold bifurcation --
\textit{i.e.}, a first order phase transition for $Y<1$ that ends
in the critical point $(g_c,Y_c)$.
(ii) a bifurcation to cycle-2
that separates SR from AR when $\rho^+=1/2$, because the refractory
period does not allow a stable fixed point
with $\rho^+ > 1/2$, generating bursts of synchronized
activity with period 2~ms.
(iii) a flip bifurcation at 
$g_{\rm Flip} =  p/q + 1/(q\Gamma J)$
that separates the uniform AI 
from the oscillatory SI in the low activity regime.
(iv) the line $Y_c=1$ is a
continuous transcritical bifurcation
for $g>g_c$ and $g<g_{\rm Flip}$;
and a synchronization phase transition for $g>g_{\rm Flip}$ 
(Fig.~\ref{PHI}c, inset).

The critical balanced point at $(g_c,Y_c)$ displays
power-law distributed avalanches with 
exponents $\tau=1.5$ and $\tau_t=2$
for size and duration, respectively, see Fig.~\ref{aval}b. 
The avalanches also
respect the scaling law $1/(\sigma\nu z)
=(\tau_t-1)/(\tau-1)$ (inset in Fig.~\ref{aval}b),
as expected for the DP universality
class~\citep{Dickman1999,Girardi2016b,Girardi2018},
and observed  in experiments~\citep{Beggs2003}. 

The simulated network activity in all the 
six dynamical regimes is shown in Fig.~\ref{aval}c.
The critical point ($g_c = 3.5, Y = 1$) 
displays avalanches sparked by a
vanishing external stimulus.
The self-sustained activity regime ($g < g_c$), 
when summed up to an
external current $Y>1$, generates the regular 
microscopic behaviors, SR or AR.
The SR state is a marginally stable 
cycle-2 of the firing density and the AR
is a state of high and homogeneous activity 
[the $\rho^+$ in Eq.~\eqref{FOT}].
The addition of an external 
current to the inhibition dominated
quiescent regime results in the 
low irregular activity AI   
(and SI if $g > g_{\rm Flip}$).
Slow oscillations (SO) are observed 
when $Y \gtrsim Y_c=1-\mu$
and $g\geq g_c$ for $\mu\geq0$.

\begin{figure}[tp]
\begin{center}
\centerline{
\includegraphics[width=0.6\textwidth]{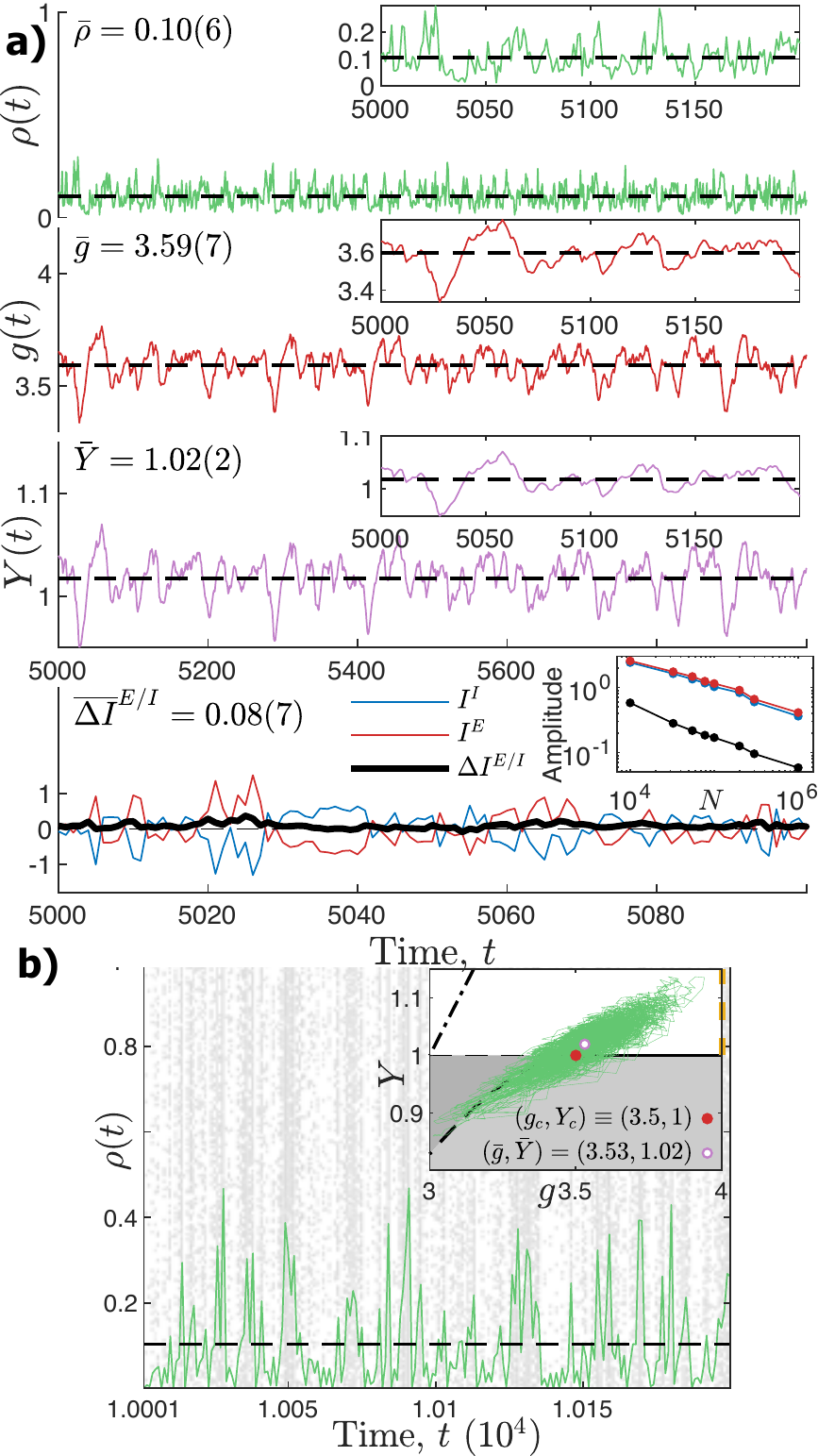}}
\caption{\label{FIG10}{\bf Self-organization towards the balanced
critical point.} Parameters: $\tau_W=\tau_\theta=100$, $A = 73.5$,
$u_W=u_\theta=0.1$, $\Gamma=1$ and $J = 10$. 
{\bf a,} Time series for $\rho[t]$, $g[t]=W^{II/EI}[t]/J$,
$Y[t] = I/\theta[t]$, and the synaptic currents
with $\overline{\Delta I}^{E/I}=0.08(7)$. $I^E$ and $I^I$ have been displaced by
their means. The amplitude of $I^E$ and $I^I$ are one
order of magnitude larger than $\Delta I^{E/I}$ for all $N$ (bottom
inset).
{\bf b,} Detail of the SOqC $\rho[t]$ dynamics
with a raster plot of 1,000 randomly
selected neurons displaying AI-like activity.
{\bf b (inset),} Self-organization trajectories in
the $g$ \textit{vs.} $Y$ plane.
The system hovers around the critical balanced point of the static model,
$g_c=3.5$ ($\bar{g}=3.59(7)$) and $Y_c=1$ ($\bar{Y}=1.02(2)$),
which displays power-law avalanches.}
\end{center}
\end{figure}


\section{Homeostatic SOqC dynamics} 
The dynamics in the inhibitory weights tunes
the system along the $g$ axis of
the phase diagram.
Threshold adaptation regulates the system along the $Y$ axis.
Both mechanisms contribute to self-organize the
network towards the critical point. 
For the parameters considered in Fig.~\ref{aval}a,
the critical point is $g_c=3.5$ and $Y_c=1$, 
and the two independent
dynamics yield
$\bar{g} = \overline{\avg{g_{ij}[t]}}=3.59(7)$ and
$\bar{Y} = \overline{\avg{Y_i[t]}}=1.02(2)$ 
(Fig.~\ref{FIG10}a). 

\begin{figure}[tp]
\begin{center}
\centerline{
\includegraphics[width=0.5\textwidth]{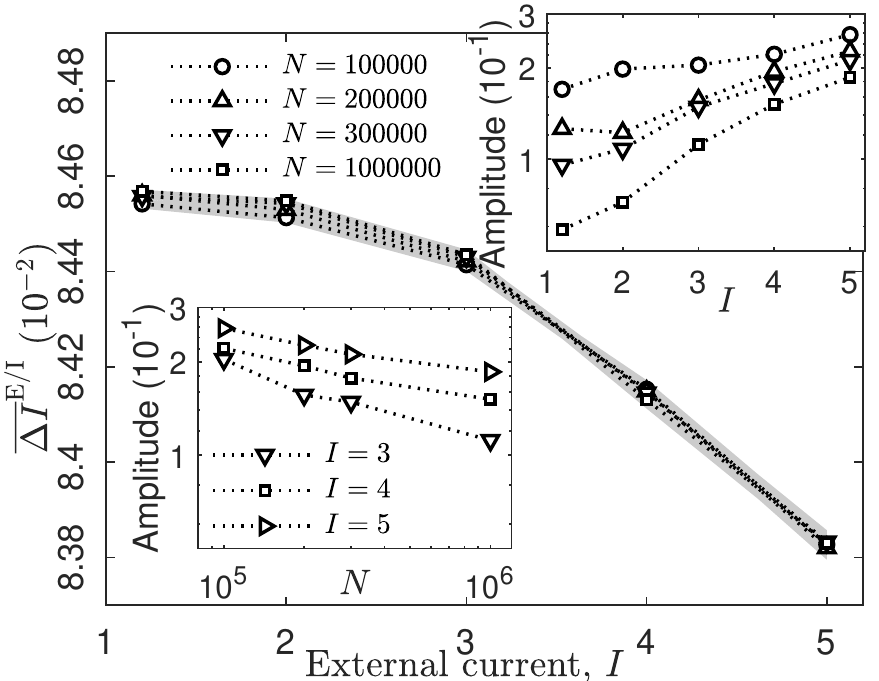}}
\caption{{\bf Synaptic balance in SOqC.}
Net synaptic current $\overline{\Delta I}^{E/I}$
as function of external input $I$.
The mean net synaptic current decreases with increasing $I$, making
the network more balanced.
Insets: Amplitude of the fluctuations of $\Delta I^{E/I}$ for different
intensities of the external input (top) and for increasing network
size (bottom). The decrease of the amplitude with
increasing $N$ shows that the fluctuations in $\Delta I^{E/I}$ 
are due to finite-size effects.}
\label{syncurr}
\end{center}
\end{figure}

This homeostatic tuning, however,
is not perfect, since stochastic oscillations 
make the system hover around the critical point 
-- a distinctive feature of 
Self-Organized quasi-Criticality or
SOqC~\citep{Bonachela2009,Bonachela2010} -- 
see Fig.~\ref{FIG10}b inset.
This oscillation is triggered by
finite-size (demographic) noise and its
amplitude decreases with increasing $N$ (inset in the bottom panel of
Fig.~\ref{FIG10}a and bottom inset of Fig.~\ref{syncurr}).
Thus, the larger the network, the closer
the system gets to the critical point~\cite{Kinouchi2019}.

The spiking pattern
of the SOqC dynamics is very similar to
standard AI activity (compare Fig.~\ref{FIG10}b 
with Fig\ref{aval}c).
This happens because
the E/I synaptic currents 
(defined in Eq.~\eqref{netsyn}) cancel
each other in fast time scales,
generating a net current
$\Delta I^{E/I}$ that is always one order of magnitude smaller than
either $I^E$ or $I^I$ (bottom panel in Fig.~\ref{FIG10}).

Contrary to the static version of the model, increasing the external input $I$ on the
homeostatic system slightly decreases the average
net current, but increases the fluctuations of $\Delta I^{E/I}$ (see Fig.~\ref{syncurr}):
the network gets more balanced and more noisy at the same time.
On the other hand, independently of $I$, the fluctuations of the net synaptic current
decreases with $N$ due to finite-size effects.
%

The nearly total cancellation of E/I currents
generates sporadic fluctuations of activity the spread throughout the network
(the avalanches) in an AI fashion. These avalanches should 
converge to nearly perfect power-law distributions for
large enough $\tau_W=\tau_\theta$~\cite{Costa2017}.
Such stochastic oscillations should have low amplitude~\cite{Kinouchi2019},
but rare large events (dragon-kings) also occur~\cite{Costa2017}.
Although the demographic noise vanishes in the thermodynamic limit,
other sources of biological noise (not included in the model and 
that does not vanishes for large $N$)
will continue to trigger the stochastic 
oscillations and the AI behavior.



\section{Discussion}
In contrast to our model, Brunel~\citep{Brunel2000} used a random network, 
deterministic LIF neurons, noisy inputs and 
a distribution of delays in the synapses.
In our model, noise is captured by the intrinsic stochasticity
of the neurons. Our model does not have
a distribution of synaptic delays, but its discrete time step implies
that spikes are transmitted with a fixed delay of 1~ms.
Also, since $\Phi(0)=0$, the reset of 
voltages after spiking implements a refractory
period of 1~ms. The other ingredients
do not seem to be crucial to obtain either
the synchronicity/activity states 
or the critical balanced point.

Our mean-field calculation is valid for fully
connected networks where the
number of neighbors is $K=N-1$.
When there is no threshold $\theta$ nor external current $I$,
the condition $W=(p-qg)J$
allows our model to be directly mapped on the
Kinouchi \textit{et al.}~\citep{Kinouchi2019} model.
In turn, the authors showed that the latter model presents
exactly the same dynamics as the sparse random 
network of probabilistic cellular automata
where $K={\cal O}(1)$, both in the static
version~\citep{Kinouchi2006,Wang2017} and
in the homeostatic version~\citep{Costa2015}.
All these models share the mean-field DP results obtained
here~\citep{Brochini2016}.
Calculations for the case $K = {\cal O}(\sqrt{N})$, 
as studied in~\citep{Van1996,Renart2010}, should be 
done to check the performance of the homeostatic mechanisms.

Heavy-tailed synaptic distributions are also expected to generate
a critical point for threshold neurons~\cite{Kusmierz2019}.
Our mean-field calculations do not apply directly in this context,
but our homeostatic mechanisms could still be employed to synaptic weights
and thresholds to check
whether the critical point would also become an attractor of that model.

While inhibition frequently increases together with excitation
after the stimulation of a neuron, the reverse
does not seem to happen; that is, 
excitation does not compensate for
inhibition when the neuron is suppressed~\citep{Lambert1994,Deneve2016,Ma2019}.
This suggests a self-organizing 
homeostatic mechanism regulating the
inhibitory synapses, which was suggested to be necessary to
re-establish power-law neuronal avalanches
in rats~\citep{Ma2019}.

This fact motivated the addition 
of adaptation to our model.
We showed that two homeostatic  mechanisms are sufficient
to take the network towards any critical balance point.
Adding homeostasis, we avoided
fine tuning of the $g$ and $Y$ parameters towards
$g_c$ and $Y_c$. However, that comes at the cost of
introducing five new parameters ($A$, $\tau_W$, $u_W$, $\tau_\theta$, $u_\theta$)
that perhaps should also be fine tuned.
This is not the case: the dependence 
on these parameters is weak,
representing a kind of gross tuning~\citep{Costa2015,Costa2017}.
Also, if necessary, 
metaplasticity in longer time scales can be employed to
tune these homeostatic parameters~\citep{Peng2013}.

\section{Concluding remarks} 
Homeostatic adaptation 
in synapses and firing thresholds are
sufficient mechanisms to 
self-organize a neuronal network
towards its DP critical (and synaptically balanced) point.
The hovering around this attractor
is due to small fluctuations in the
net synaptic current, such that
there is always some residual excitation driving the network
activity (a sort of fluctuation-driven AI regime due to SOqC).
The underlying critical point
shows power-law avalanches with
exponents compatible with \textit{in vitro} 
experiments~\citep{Beggs2003}.
Our model thus unifies two different perspectives on the spontaneous activity of the brain:
power-law neuronal avalanches and fluctuation
driven asynchronous-irregular firing 
patterns are indeed two sides
of the same coin.

\begin{acknowledgments}
We thank A. C. Roque, M. Copelli and J. Stolfi for discussions. 
This article was produced as part of the
FAPESP Research, Innovation and Dissemination Center
for Neuromathematics (grant \#2013/07699-0,
S. Paulo Research Foundation).
L.B. thanks FAPESP (grant \#2016/24676-1), 
A.A.C. thanks FAPESP (grants \#2016/00430-3 and 
\#2016/20945-8) and M.G.-S. thanks FAPESP (grant 
\#2018/09150-9).
O.K. thanks the Center for Natural and
Artificial Information Processing Systems (CNAIPS)-USP 
and FAPESP BPE grant \#2019/12746-3.
The present work was also realized with the support of
CNPq, Conselho Nacional de Desenvolvimento Cient{\'i}fico  
e Tecnol{\'o}gico, Brazil.
\end{acknowledgments}

\bibliographystyle{unsrt}

\end{document}